\documentclass[showpacs,aps,graphicx,twocolumn]{revtex4}
\usepackage{graphicx}

\begin{document}

\title{Residual effect on the robustness of multiqubit  entanglement\footnote{published in Phys. Rev. A \textbf{82}, 014301 (2010)}}
\author{Bao-Kui Zhao$^{1,2}$ and Fu-Guo Deng$^{1}$\footnote{
Corresponding author: fgdeng@bnu.edu.cn.} }
\address{
$^1$Department of Physics, Applied Optics Beijing Area Major
Laboratory, Beijing Normal University, Beijing
100875, China\\
$^2$College of Nuclear Science and Technology, Beijing Normal
University, Beijing 100875,  China}
\date{\today }

\begin{abstract}
We investigate the relation between the entanglement and the
robustness of a multipartite system to a depolarization noise. We
find that the robustness of a two-qubit system in an arbitrary pure
state depends completely on its entanglement. However, this is not
always true in a three-qubit system. There is a residual effect on
the robustness of a three-qubit system in an arbitrary superposition
of Greenberger-Horne-Zeilinger state and W state. Its entanglement
determines the trend of its robustness. However, there is a
splitting on its robustness under the same entanglement. Its
robustness not only has the same periodicity as its three-tangle but
also  alters with its three-tangle synchronously. There is also a
splitting on the robustness of an $n$-qubit ($n>3$) system although
it is more complicated.
\end{abstract}
 \maketitle

Entanglement is the most nonclassical feature of quantum mechanics
and it plays an important role in quantum computation and
communication \cite{Nielsen}. However, an entangled quantum system
is inevitably immersed in an environment and interacts with it in
some way, which usually degrades the entanglement of the system.
This decoherence introduces some disadvantages on multiqubit
entanglement creation and manipulation in quantum information
processing. Recently, some works showed that entanglement sudden
death (ESD) \cite{Dodd,Yu}, a peculiar dynamical feature of
entangled states, may takes place in an entangled system. In detail,
the entanglement of an entangled system may disappear at a finite
time although the constituent parts of an entangled state decay
asymptotically in time. This interesting phenomenon has been
observed in a two-qubit optical system \cite{ESDexp,ESDexp2}.

Recently, some groups have studied the robustness of mulitqubit
quantum systems. For example, in 1999, Vidal and Tarrach
\cite{Vidal} investigated the robustness of two-qubit systems by
considering the minimal amount of mixing with locally prepared
states which washes out all entanglement. In 2002, Simon and Kempe
\cite{Simon} showed that the robustness of
Greenberger-Horne-Zeilinger (GHZ) entanglement increases with the
number of qubits, under local decoherence, modeled by partially
depolarizing channels acting independently on each qubit. They also
pointed out that the amount of entanglement of a multiqubit system
and its robustness do not have a simple relation. In 2008, Aolita
\emph{et al.} \cite{Aolita} showed clearly that  the time at which
such entanglement of a pure GHZ-state quantum system becomes
arbitrarily small is inversely proportional to $N$, although its ESD
time increases with $N$. In 2008, Batle and Casas \cite{ijqi} found
that the entanglement in a three-qubit system with the measure of
Mermin's inequality \cite{Mermin} and its robustness contain
similarities, but do not have a simple relation.  In 2009, Liu and
Fan \cite{Liu} investigated the decay of entanglement of a
generalized $N$-qudit GHZ state under local decoherence and obtained
results similar to those reported in Refs.\cite{Simon,Aolita}. In
2009, Borras \emph{et al.} \cite{Borras} investigated the decay of
the amount of entanglement of a multqubit system experiencing a
decoherence process.

In this article, we investigate the relation between the
entanglement and the robustness of a multipartite system to a
depolarization noise. For a two-qubit system in an arbitrary pure
entangled state, its robustness increases synchronously with its
entanglement. For a three-qubit system, this phenomenon, however,
disappears and the robustness of GHZ-like states, which have only
three-tangle (the entanglement shared by all the three qubits)
without concurrence, is the upper boundary of symmetrical
three-qubit pure states. It is interesting to point out that there
is a residual effect on the robustness of a three-qubit system in an
arbitrary superposition of a GHZ state and a W state. Its
entanglement, on the one hand, determines the trend of its
robustness. On the other hand, there is a splitting on its
robustness. Its robustness not only has the same periodicity as its
three-tangle but also alters with its three-tangle synchronously.
This interesting phenomenon takes place in a four-qubit system and a
five-qubit system, but there is not a good measure on their
entanglements shared by all the qubits in this time.

In the present study, each qubit in multiqubit entangled quantum
systems is coupled to its own environment individually. That is, our
study is under local decoherence, modeled by partially depolarizing
channels acting independently on each qubit, the same as that in
Ref.\cite{Simon}. The dynamics of each qubit in a three-particle
entangled quantum system is governed by a master equation from which
one can obtain a completely positive trace-preserving map
$\varepsilon$ which describes the corresponding evolution
\cite{Borras}: $\rho_{i}(t)=\varepsilon(t)\rho_{i}(0)$. In the
Born-Markovian approximation the maps (or channels) can be described
using its Kraus representation \cite{Borras,booknoise}, that is,
\begin{eqnarray}
\varepsilon(\rho_{i}(0))=\sum^3_{j=1}E_{ji}(t)\rho_{i}(0)E_{ji}^{\dag}(t),
\end{eqnarray}
where $E_{j}(t)$ ($j=1,2,3$) are the so-called Kraus operators
needed to completely characterize the channel. In detail, the
partially depolarizing channel $C_{d}$ for each qubit is defined by
applying the completely depolarizing channel with a probability $d$,
and applying the identity transformation with a probability $1-d$.
This corresponds to the following transformation \cite{Simon}
\begin{eqnarray}
|i\rangle \langle j| & \longrightarrow & (1-d)|i\rangle \langle
j|+d\delta_{ij}\frac{1}{2}\textbf{1}\label{noise}\\
P_{k}& \longrightarrow &\frac{1+s}{2}P_{k}+\frac{1-s}{2}P_{k\oplus 1}\\
\sigma_{+}& \longrightarrow & s\sigma_{+}\\
\sigma_{-}& \longrightarrow & s\sigma_{-}.
\end{eqnarray}
Here $s=1-d$, $k\in\{0,1\}$, and $k\oplus 1$ means that  the sum of
$k$ and $1$ mod 2. $P_0=\vert 0\rangle\langle 0\vert$, $P_0=\vert
1\rangle\langle 1\vert$, $\sigma_+=\vert 0\rangle\langle 1\vert$,
and $\sigma_-=\vert 1 \rangle\langle 0\vert$.

First we explore the connection between the entanglement of a
two-qubit quantum system and its robustness and then generalize it
in a three-qubit system.

\begin{figure}[!h]
\begin{center}
\includegraphics[width=8cm,angle=0]{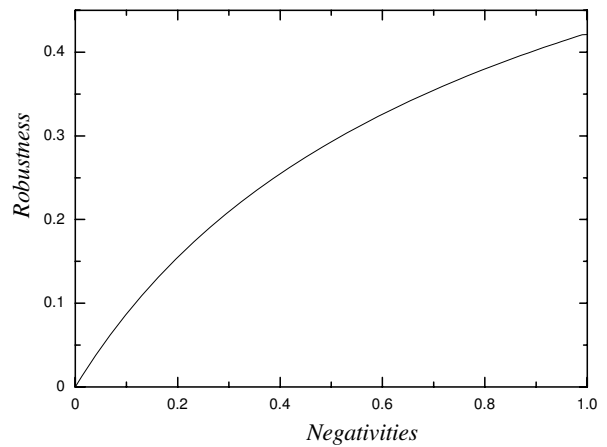}
\caption{The relation between the  robustness and the negativity of
a two-qubit system in a pure entangled state.}\label{fig1}
\end{center}
\end{figure}

Under the depolarized noise $C_d$ shown in Eq(\ref{noise}) on each
qubit, the robustness of a given $n$-party entangled state $\rho$ is
defined as the critical amount of depolarization $d_{crit}$ where
$C_d^{\otimes n}(\rho)$ becomes separable, or-- in the absence of a
necessary and sufficient condition-- ceases to fulfill certain
sufficient conditions for entanglement
\cite{Simon,depcon2,depcon3,depcon1}. The entanglement
(negativity--$N$)  of a quantum system in an arbitrary pure state
can be obtained with the method in
Refs.\cite{Negavity1,Negavity2,Negavity3} and its robustness
($R=d_{crit}$) can be calculated with the PPT criterion
\cite{depcon2}. Negativity of a bipartite quantum system is defined
as $N=\Sigma \vert \lambda \vert$, where $\lambda$ are the negative
eigenvalues of $\rho^{T_B}$ and $T_B$ denotes the partial transpose
with respect to the subsystem $B$. For an arbitrary bipartite pure
state, its entanglement is completely determined by its Schmidt
coefficients, and the depolarizing channel is basis independent.
With a numerical simulation, we give the relation between the
entanglement and the robustness in two-qubit pure states in
Fig.\ref{fig1}. It is explicit that the more the entanglement of a
two-qubit entangled pure state, the more its robustness.
 That is,
the two systems with the same entanglement have the same robustness
although they are in two different entangled pure states. Also, we
find that there is a simple analytical expression for the relation
between $N$ and $R$ of  a two-qubit system in an arbitrary pure
state, that is,
\begin{eqnarray}
R=1-\frac{1}{\sqrt{1+2N}}.
\end{eqnarray}
That is, its robustness  depends completely on its entanglement.

For a three-qubit system, there are two inequivalent classes of
genuine tripartite entanglement \cite{Dur}, represented by the GHZ
state $|GHZ\rangle = \frac{1}{\sqrt{2}}(|000\rangle + |111\rangle)$
and the W state $|W\rangle  = \frac{1}{\sqrt{3}}(|001\rangle +
|010\rangle + |100\rangle)$. Each three-qubit pure state can be
converted into either a GHZ-class state or a W-class state by
stochastic local quantum operations assisted by classical
communication (SLOCC) \cite{Bennett,Dur}. Our question is, whether
or not a three-qubit quantum system has the same result as a
two-qubit quantum system? That is, is there the explicit phenomenon
that the more the entanglement of a three-qubit entangled pure
state, the more its robustness?

\begin{figure}[!h]
\begin{center}
\includegraphics[width=8cm,angle=0]{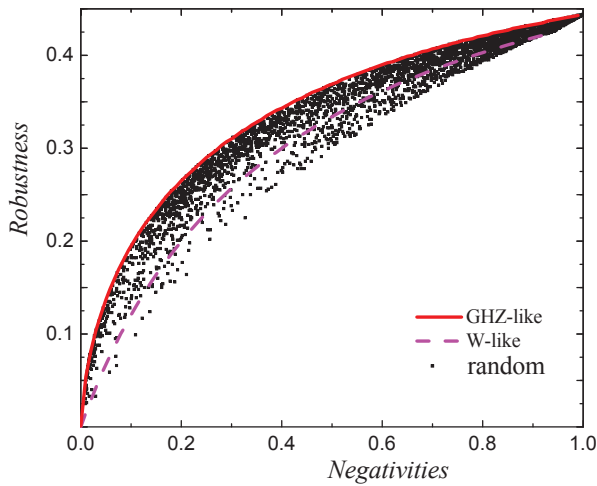}
\caption{(Color online) The relation between the  robustness and the
negativity of a three-qubit system in a GHZ-like state, a W-like
state,  and a symmetrical three-qubit pure state is shown with a
solid (red) line, a  dashed (magenta) line,  or a solid (black) dot,
respectively. For a GHZ-like state, negativity $\in[0,1]$. For a
W-like state, negativity
$\in[0,\frac{2\sqrt{2}}{3}]$.}\label{fig_robustness}
\end{center}
\end{figure}

As the depolarizing channel is symmetric to each qubit in a
three-qubit system,  we consider two classes of symmetric pure
states, which are invariable under the permutations of the three
particles, to study the relation between their robustness and  their
entanglements, i.e., the GHZ-like state $|\Lambda\rangle$ and the
W-like state $|\Omega\rangle$,
\begin{eqnarray}
|\Lambda\rangle_{ABC} &=& \sqrt{a}|000\rangle+\sqrt{1-a}|111\rangle,
\nonumber\\
|\Omega\rangle_{ABC} &
=&\sqrt{b}|000\rangle+\sqrt{\frac{1-b}{3}}(|001\rangle+|010\rangle+|100\rangle),\nonumber
\end{eqnarray}
where $a,b\in[0,1]$. The relations between the entanglements and the
robustness of the states $|\Lambda\rangle$ and $|\Omega\rangle$ are
plotted with the solid (red ) line and the  dashed (magenta) line,
respectively, in Fig.\ref{fig_robustness}. For each class of
entangled states, on the one hand, the more the entanglement, the
more its robustness. On the other hand, the GHZ-like state
$|\Lambda\rangle_{ABC}$ is more robust than the W-like state
$|\Omega\rangle_{ABC}$  with the same entanglement under the
depolarizing channel $C_d^{\otimes 3}$.

More generally, an arbitrary symmetrical three-qubit entangled pure
state can be written as
\begin{eqnarray}
|\Psi\rangle_{ABC}=c_1 |000\rangle + c_2 \frac{1}{\sqrt{3}}\{|001\rangle+|010\rangle+|100\rangle\}\nonumber\\
+ c_3 \frac{1}{\sqrt{3}}\{|110\rangle+|101\rangle+|011\rangle\} +
c_4|111\rangle,
\end{eqnarray}
where $c_1$, $c_2$, $c_3$, and $c_4$ are four complex numbers and
satisfy $|c_1|^2+|c_2|^2+|c_3|^2+|c_4|^2=1$. We choose randomly a
great number of  symmetrical three-qubit entangled pure states with
the Haar measure \cite{Zyczkowski} and calculate their negativity.
With a numerical simulation, we give the relation between the
entanglement and the robustness in symmetrical three-qubit pure
states $|\Psi\rangle_{ABC}$ in Fig.\ref{fig_robustness} with solid
black dots. Different from two-qubit states, the robustness of a
symmetrical three-qubit state does not only depend on its
negativity. Generally speaking, the more the entanglement, the more
the robustness of a symmetrical three-qubit pure state.  However,
one can see that there is a subset of states which have the same
entanglement but different robustness. Moreover, the GHZ-like state
is more robust than others under the same entanglement. That is, the
robustness of GHZ-like states, which have only three-tangle
\cite{CKW,threetangle,Lohmayer} (it is called originally the
residual entanglement \cite{CKW}) without concurrence, is the upper
boundary of symmetrical three-qubit pure states, but  W-like states
without three-tangle are not the most fragile ones. This phenomenon
implies that the three-tangle is the highest-quality (the most
robust) entanglement against the depolarization noise.

\begin{figure}[!h]
\begin{center}
\includegraphics[width=8cm,angle=0]{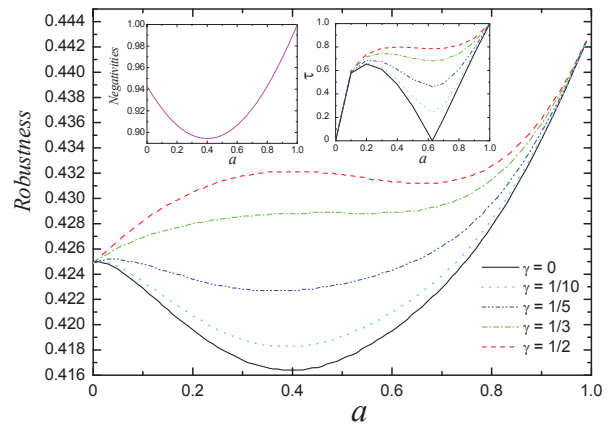}
\caption{(Color online) Robustness of the state $\vert
Z(a,\varphi)\rangle=\sqrt{a}|GHZ\rangle-e^{i\varphi}\sqrt{1-a}|W\rangle$
versus the coefficient $a$ with different phase facotrs
$\varphi=\gamma\cdot\frac{2\pi}{3}$ (rad) (from top to bottom:
$\gamma=1/2, 1/3, 1/5, 1/10, 0$). The left inset shows the relation
between the entanglement of the state $\vert Z (a,\varphi)\rangle$
and the coefficient $a$. The right inset shows the residual
entanglement of the state $\vert Z (a,\varphi)\rangle$ versus the
coefficient $a$ with different phase factors
$\varphi$.}\label{fig_robustness2}
\end{center}
\end{figure}

In a three-qubit system, the negativity only measures a bipartite
entanglement, while the three-tangle quantifies genuine multipartite
correlations. In order to explore the role of different entanglement
components (three-tangle or negativity) in robustness of a
three-qubit system, we should study the robustness of a class of
three-qubit pure states which have the same entanglement and
different robustness. Fortunately, we find that the superpositions
of the GHZ state and the W state have different robustness under the
same entanglement (negativity), that is,
\begin{eqnarray}
\vert Z (a,\varphi)\rangle&=&
\sqrt{a}|GHZ\rangle-e^{i\varphi}\sqrt{1-a}|W\rangle.
\end{eqnarray}
The parameter $a\in[0,1]$ is used to identify the proportion of the
GHZ state and $\varphi$ is used to represent the relative phase
between the GHZ state and the W state. The negativities of these
three-qubit systems can be described as
\begin{eqnarray}
N[\vert Z(a,\varphi)\rangle]&=&\frac{\sqrt{5a^{2}-4a+8}}{3}.
\end{eqnarray}
It is obvious that the entanglement of the state $\vert Z
(a,\varphi)\rangle $ does not depend on the phase factor $\varphi$.
The relation between its entanglement and the coefficient $a$ is
shown in the left inset in Fig.\ref{fig_robustness2}. However, its
robustness depends on not only the coefficient $a$ but also the
phase factor $\varphi$, shown in Fig.\ref{fig_robustness2} for the
cases $\varphi=\pi/3,2\pi/9, 2\pi/15, \pi/15, 0$. Moreover, its
robustness $R[\vert Z(a,\varphi)\rangle]$ varies periodically with
the phase $\varphi$; that is,
\begin{eqnarray}
R[\vert Z(a,\varphi)\rangle]=R[\vert
Z(a,\varphi+2\pi/3)\rangle].\nonumber
\end{eqnarray}
The periods of the robustness $T=2\pi/3$. In each period,
\begin{eqnarray}
R[\vert Z(a,\pi/3-\Delta\varphi+nT)\rangle]=R[\vert
Z(a,\pi/3+\Delta\varphi+nT)\rangle],\nonumber
\end{eqnarray}
where $\Delta\varphi\in [0,\pi/3]$ and $n=0,1,2,\ldots$.

With a given coefficient $a$, which means that the states $\vert
Z(a,\varphi)\rangle$ have the same components of the GHZ state and
the W state, the larger the phase $\varphi$, the greater the
robustness of the state $\vert Z(a,\varphi)\rangle$ in each period.
With a given phase factor $\varphi$, the robustness of the state
$\vert Z(a,\varphi)\rangle$ appears to have two different effects.
When $\varphi$ is small, a tripartite quantum system in the state
$\vert Z(a,\varphi)\rangle$ appears to have a normal robustness
effect in which its robustness alters synchronously with its
negativity. That is, its robustness decreases with the coefficient
$a$ when $a\leq 0.4$ and increases with $a$ when $a > 0.4$, same as
its negativity. However, an abnormal robustness effect takes place
when $\varphi$ becomes large. Specifically, the robustness of the
state $\vert Z(a,\varphi)\rangle$ alters inversely with its
negativity when $a$ is not large. In principle, it is explicit that
there is a splitting on the robustness of the state $\vert
Z(a,\varphi)\rangle$, which is completely different from its
negativity. The fluctuation of the robustness is at most 4\%. That
is, the negativity of a three-qubit system determines the trend of
its robustness. Also, there is a residual effect on the robustness
although the fluctuation is small.

It is interesting to point out that the feature of the robustness of
the state $\vert Z(a,\varphi)\rangle$ agrees with that of  its
three-tangle. For the state $\vert Z(a,\varphi)\rangle$, its
three-tangle  can be calculated as \cite{Lohmayer}
\begin{eqnarray}
\tau[\vert
Z(a,\varphi)\rangle]&=&|a^{2}-\frac{8\sqrt{6}}{9}\sqrt{a(1-a)^{3}}e^{3i\varphi}|.
\end{eqnarray}
It is periodic in the phase $\varphi$ with  the same periods of
$T_\tau=T=2\pi/3$ as the robustness of the sate  $\vert
Z(a,\varphi)\rangle$, shown in the right insert in
Fig.\ref{fig_robustness2} for the cases $\varphi=\pi/3,2\pi/9,
2\pi/15, \pi/15,0$. Also, in each period,
\begin{eqnarray}
\tau[\vert Z(a,\pi/3-\Delta\varphi+nT)\rangle]=\tau[\vert
Z(a,\pi/3+\Delta\varphi+nT)\rangle].\nonumber
\end{eqnarray}

\begin{figure}[!h]
\begin{center}
\includegraphics[width=8cm,angle=0]{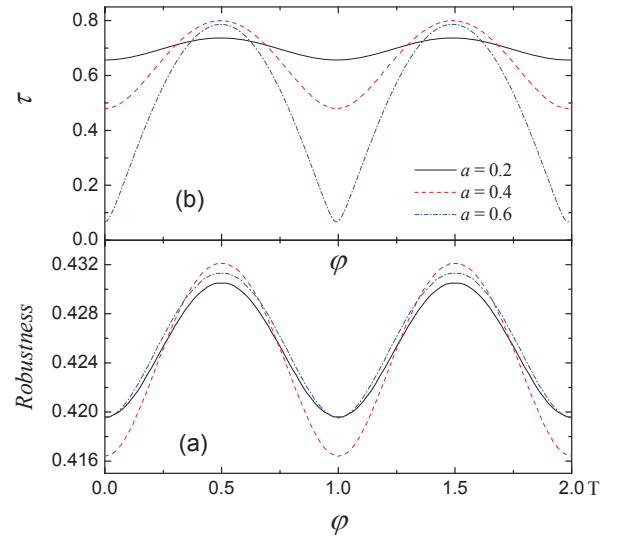}
\caption{(Color online) (a) Periodicity  and monotonicity of the
robustness of the state $\vert
Z(a,\varphi)\rangle=\sqrt{a}|GHZ\rangle-e^{i\varphi}\sqrt{1-a}|W\rangle$
versus $\varphi=\gamma\cdot\frac{2\pi}{3}$ (rad) with a different
coefficient  $a$ (from top to bottom: $a=0.2, 0.4, 0.6$); (b)
Periodicity   and monotonicity of the three-tangle of the state
$\vert Z(a,\varphi)\rangle$ versus $\varphi$ with  a different
coefficient $a$.}\label{fig_robustness3}
\end{center}
\end{figure}

In order to show explicitly the effect of the three-tangle on the
the robustness of the state $\vert Z(a,\varphi)\rangle$, we give the
relation between the robustness $R$ and the phase $\varphi$ and that
between $\tau$ and the phase $\varphi$  under a given coefficient
$a=0.2,0.4,0.6$, shown in  Fig.\ref{fig_robustness3}. For a state
$\vert Z (a,\varphi)\rangle$ under a given coefficient $a$, both its
robustness and its three-tangle $\tau$ increase with the phase
$\varphi$ when $0\leq\varphi\leq \pi/3$ and decrease with the phase
$\varphi$ when $\pi/3\leq\varphi\leq 2\pi/3$. That is, the
robustness and the three-tangle  have the same monotonicity.

For a four-qubit system and a five-qubit system in the state $\vert
Z (a,\varphi)\rangle_n = \sqrt{a}|GHZ\rangle_n
-e^{i\varphi}\sqrt{1-a}|W\rangle_n$ ($n=4,5$), there is also the
phenomenon as a three-qubit system. That is, there is a splitting on
its robustness. Here $|GHZ\rangle_n=\frac{1}{\sqrt{2}}(\vert
00\cdots 0\rangle + \vert 11 \cdots 1\rangle)$ and
$|W\rangle_n=\frac{1}{\sqrt{n}}(\vert 0\cdots 01\rangle + \vert
0\cdots 10\rangle + \cdots + \vert 1\cdots 00\rangle)$. The
robustness has the periods $T_{R_n}=2\pi/n$, i.e., $R_n[\vert
Z(a,\varphi)\rangle_n]=R_n[\vert Z(a,\varphi+2\pi/n)\rangle_n]$.
Unfortunately, there is lack of studies on  $n$-tangle ($n\geq 4$,
i.e., the entanglement shared by all the $n$ qubits). For two-qubit
systems, their entanglements are just the ones shared by both the
qubits, and the robustness of each two-qubit pure state depends
completely on its negativity.

In summary, we have investigated the relation between the
entanglement and the robustness of a multipartite system to a
depolarized noise. For a two-qubit system in an arbitrary pure
entangled state, its negativity determines completely its
robustness. For a three-qubit system in a general symmetrical pure
state, the trend of its robustness is, on the one hand, determined
by its entanglement. The robustness of GHZ-like states, which have
only three-tangle without concurrence, is the upper boundary of
symmetrical three-qubit pure states, which implies that the
three-tangle is the highest-quality entanglement against the
depolarization noise.   On the other hand, there is a residual
effect on the robustness. That is, the robustness of a three-qubit
system in an arbitrary superposition of a GHZ state and a W state
not only has the same periodicity as its three-tangle but also
alters with its three-tangle synchronously. There is also a
splitting on the robustness of a four-qubit system and a five-qubit
system although there is not a good measure for their $n$-tangle
($n$=4,5).

One should certainly be careful in drawing general conclusions about
the robustness and the entanglement of a multipartite system.
Nevertheless, our results show that the three-tangle affects the
robustness of a three-qubit system as they have the same periodicity
and monotonicity but not the extent. Moreover, the splitting on the
robustness of a system with  more than three qubits provides an open
question for people to study the definition of the entanglement
shared by all the qubits. Also, it may give some useful information
about decoherence in quantum information processing. On the other
hand, are there other factors that affect the robustness of
mutli-qubit system?  Dose this phenomenon exist in high-dimensional
systems or not? These are still some open questions of interest to
us.

This work is supported by the National Natural Science Foundation of
China under Grant No. 10974020, A Foundation for the Author of
National Excellent Doctoral Dissertation of P. R. China under Grant
No. 200723, and the Beijing Natural Science Foundation under Grant
No. 1082008.

\end{document}